\documentclass[a4paper,11pt]{article}
\usepackage{pos,bm}

\title{Quasielastic Lepton-Nucleus Scattering and the Correlated Fermi Gas Model}
\author*[a]{Gil Paz}
\affiliation[a]{Department of Physics and Astronomy, \\
Wayne State University, Detroit, Michigan 48201, USA}

\emailAdd{gilpaz@wayne.edu}
\abstract{The future neutrino research program will require improved precision. A major source of uncertainty is the interaction of neutrinos with nuclei that serve as targets for such experiments. Broadly speaking, this interaction often depends, e.g., for Charge-Current Quasi-Elastic  (CCQE) scattering, on the combination of ``nucleon physics" expressed by form factors and ``nuclear physics" expressed by a nuclear model. It is important to get a good handle on both. This talk presents a fully analytic implementation of the Correlated Fermi Gas (CFG) Model for CCQE electron-nuclei and neutrino-nuclei scattering. The implementation is used to compare separately form factors and nuclear model effects for both electron-carbon and neutrino-carbon scattering data.}

\FullConference{42nd International Conference on High Energy Physics (ICHEP2024)\\
18-24 July 2024\\
Prague, Czech Republic\\}

\begin{document}
\maketitle

\paragraph{Motivation:}
Current and future neutrino experiments aim to precisely measure Standard Model parameters and uncover non-standard neutrino interactions (NSI). Precision measurements require better control of systematic uncertainties in lepton-nucleus interactions. For example, Quasiealstic $\nu-N$ scattering depends (i) on \emph{nucleon} physics via form factors and (ii) \emph{nuclear} physics via a nuclear model. 
%In other words, one is interested in the quark level process $\nu_\ell+d\to \ell^-+u$ that is folded twice. First, since the quarks are in nucleons one needs to consider $\nu_\ell + n \to \ell^- + p$, giving rise to form factors. Second, since nucleons are in nuclei, one needs to consider $\nu_\ell + \mbox{nucleus} \to \ell^- +\mbox{nucleus}^\prime$, giving rise to a nuclear model. 
Precision requires  separate control of   \emph{form factors} and \emph{nuclear effects}.

 We recently studied this question in our paper
 %a paper titled ``Quasielastic Lepton-Nucleus Scattering and the Correlated Fermi Gas Model,'' 
 \cite{Bhattacharya:2024win}. This paper has two goals. First, to analytically implement the Correlated Fermi Gas (CFG) Model of Hen, Li, Guo, Weinstein, and Piasetzky \cite{Hen:2014yfa} for lepton-nucleus scattering. Second, to use the CFG and the well-known Relativistic Fermi Gas (RFG) models to separate form factors and nuclear effects. For both models the initial (final) nucleons are distributed as $n_i({\bm p})$ ($n_f({\bm p^\prime})$), the  single nucleon cross section $\sigma_{\rm nucleon}(\bm{p} \to \bm{p}^\prime)$ depends on form factors and the nuclear cross section is schematically as  
$
\sigma_{\rm nuclear}=n_i({\bm p})\otimes\sigma_{\rm nucleon}(\bm{p} \to \bm{p}^\prime)
\otimes [1-n_f({\bm p^\prime})].
$
The following briefly describe the CFG model and compare form factor and nuclear model effects for electron-carbon and neutrino-carbon scattering data.  Because of space-time constraints this talk focuses on results in the form of plots. For a detailed discussion of the analytic expressions and a complete list of references see \cite{Bhattacharya:2024win}.

\begin{figure}
\begin{center}
\includegraphics[scale=0.23]{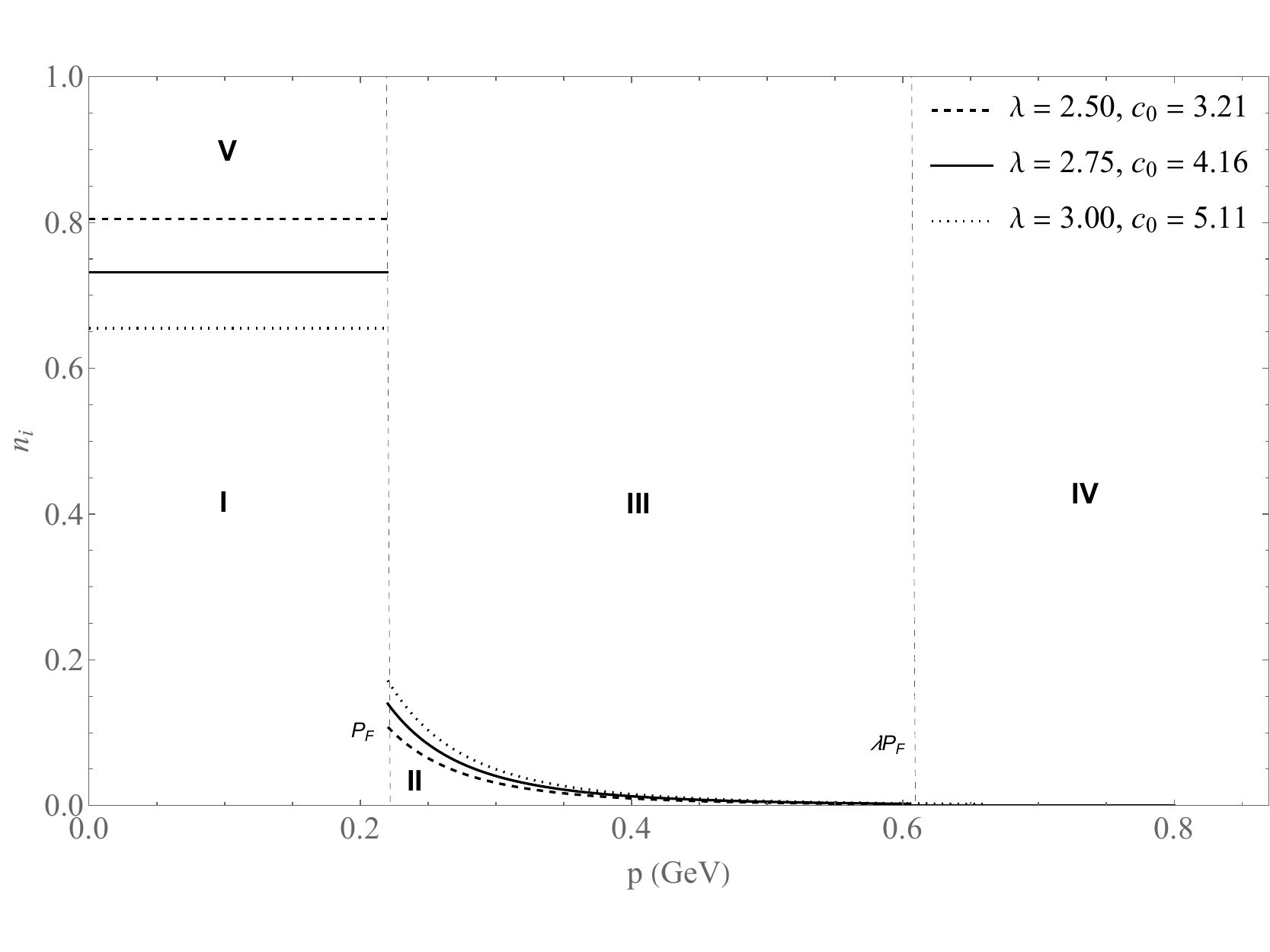}
\hspace{0.5cm}
\includegraphics[scale=0.36]{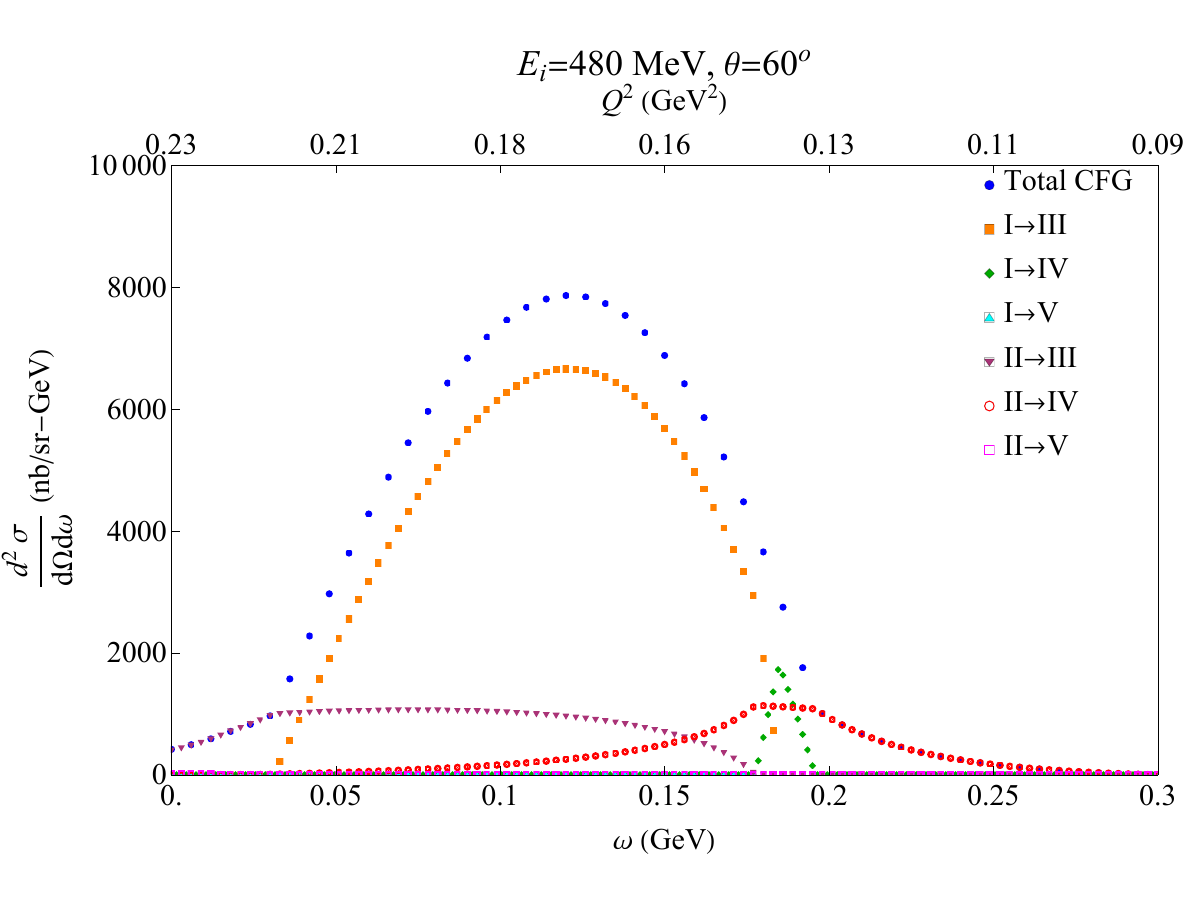}
\caption{\label{Fig:CFG} Left: $n_i (\bm{p})$ as a a function of the magnitude of the 3-momentum. The different lines correspond to extremal variation of the CFG model parameters $\lambda$ and $c_0$. Right: The contributions to the CFG model cross section by transitions. The incident electron energy is 480 MeV and the scattering angle is $60^\circ$}
\end{center}
\end{figure}

\paragraph{Analytic implementation of the CFG model:}
In the RFG model nucleons occupy states only up to the Fermi momentum. Data from the last two decades showed $\sim20\%$ of nucleons 
have momentum greater than the Fermi momentum. They  appear in short-range correlated (SRC) neutron-proton pairs. The idea of the CFG model is to ``add" high-momentum tail above the Fermi momentum \cite{Hen:2014yfa}, see figure \ref{Fig:CFG}. The initial nucleon can be in regions I or II  and the final nucleon can be in regions III, IV, or V. Therefore  there are six possible transitions.  The model includes two parameters $c_0$ and $\lambda$ with values given in \cite{Hen:2014yfa}.  In the formal limit $\lambda\to 1$ the RFG model is recovered.  Since $\lambda\approx2.75\pm0.25$, this limit is never obtained in practice.

The analytic implementation of the model is described in \cite{Bhattacharya:2024win}. Here, we just illustrate the contributions of the various transitions to the electron-carbon scattering cross section as a function of the energy transfer $\omega$ for an incident electron energy of 480 MeV and a scattering angle of $60^\circ$, see figure \ref{Fig:CFG}. The cross section is dominated by the I $\to$ III transitions, where the shape of the cross section is analogous to the RFG model. The large $\omega$ tail is generated mostly by the II $\to$ IV transition, while the small $\omega$ tail is generated by the II $\to$ III transition.

\begin{figure}
\begin{center}
\includegraphics[scale=0.35]{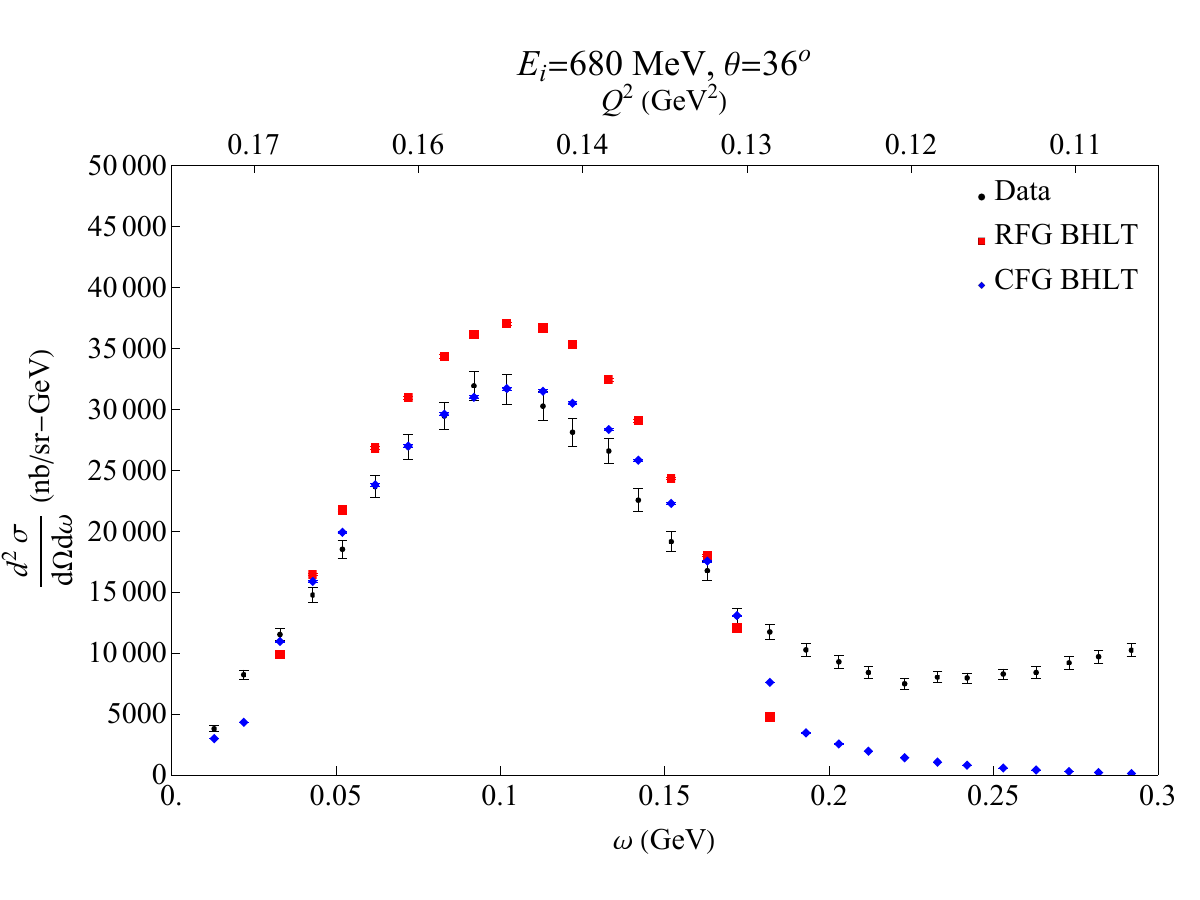}
\hspace{0.5cm}
\includegraphics[scale=0.35]{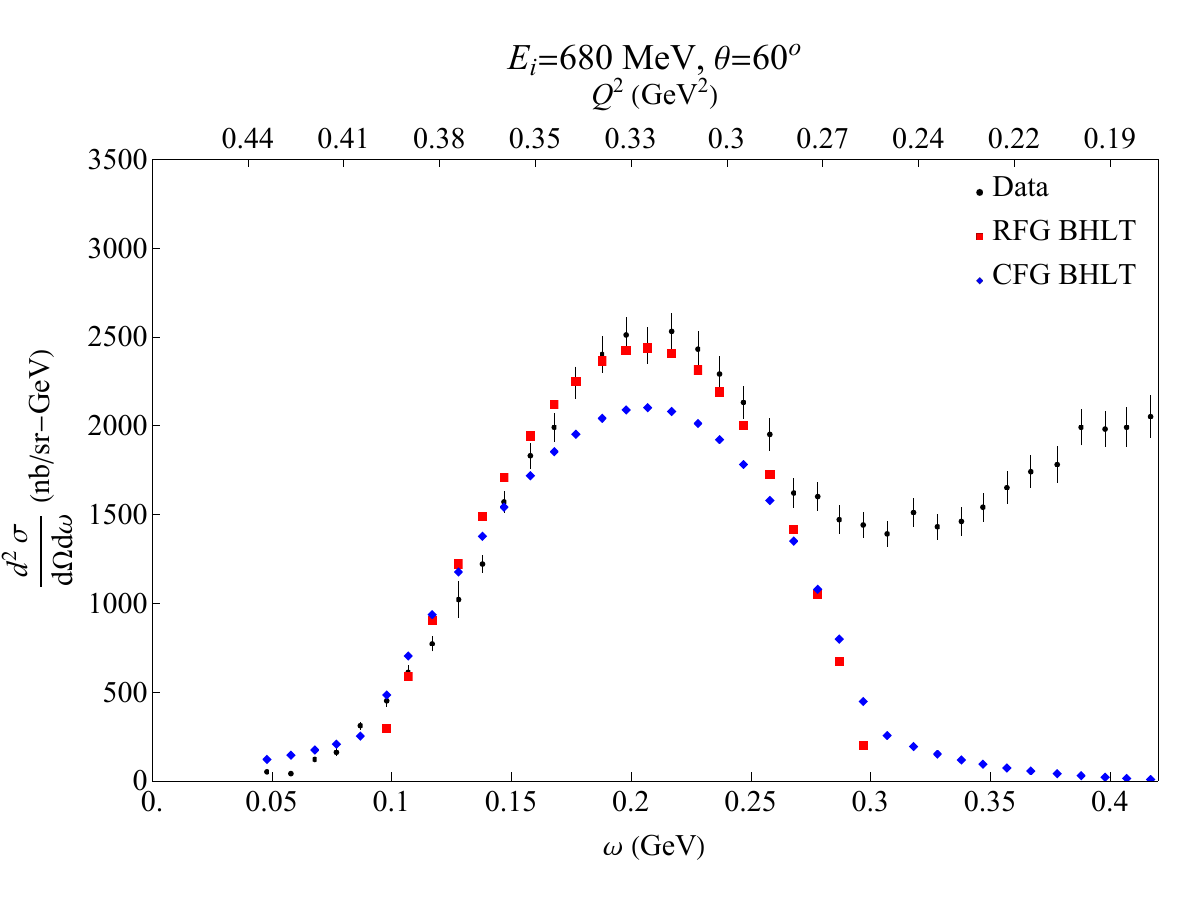}
\caption{\label{Fig:680_MeV} A comparison of RFG (red squares) and CFG (blue diamonds) models to carbon data. The incident electron energy is 680 MeV.  Left: Scattering angle of  $36^\circ$ Left: Scattering angle of  $60^\circ$. 
}
\end{center}
\end{figure}

\begin{figure}
\begin{center}
\includegraphics[scale=0.35]{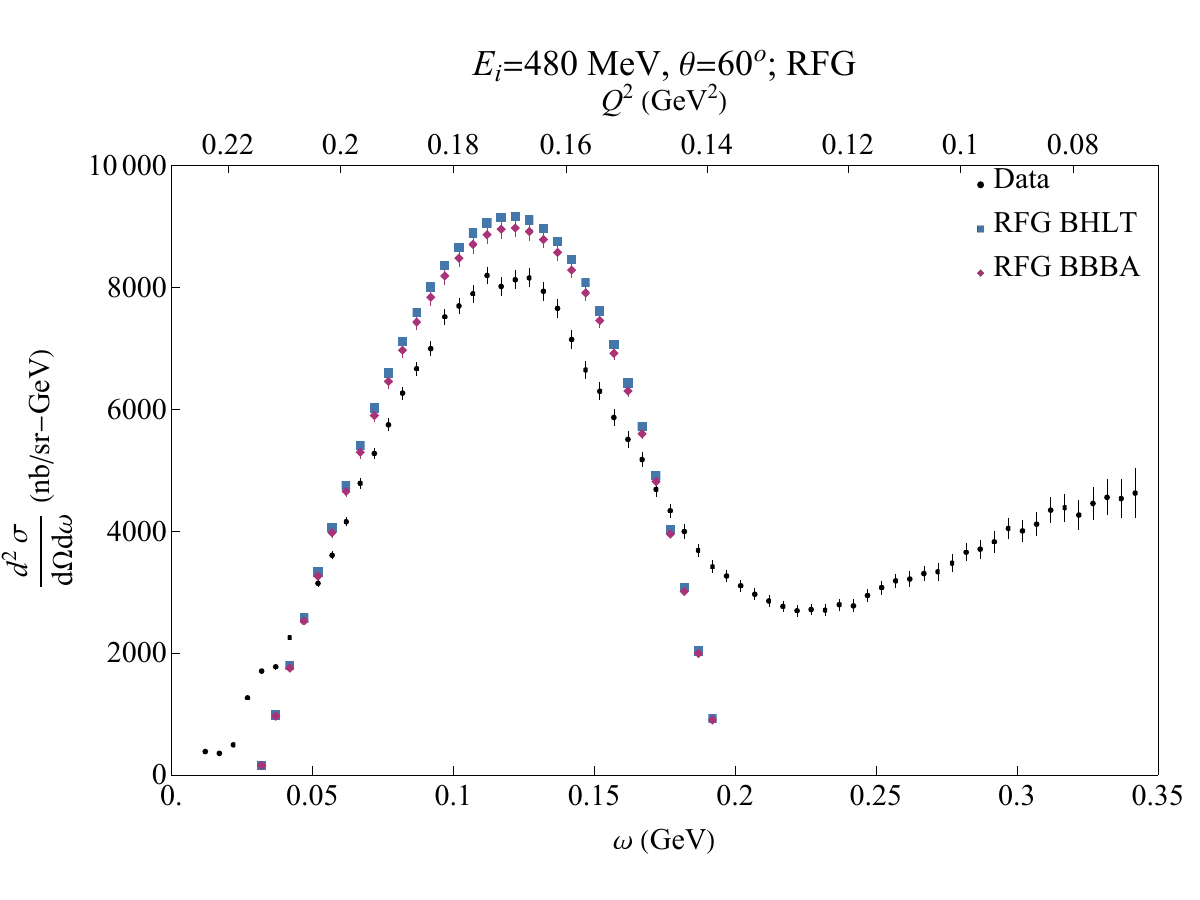}
\hspace{0.5 cm}
\includegraphics[scale=0.35]{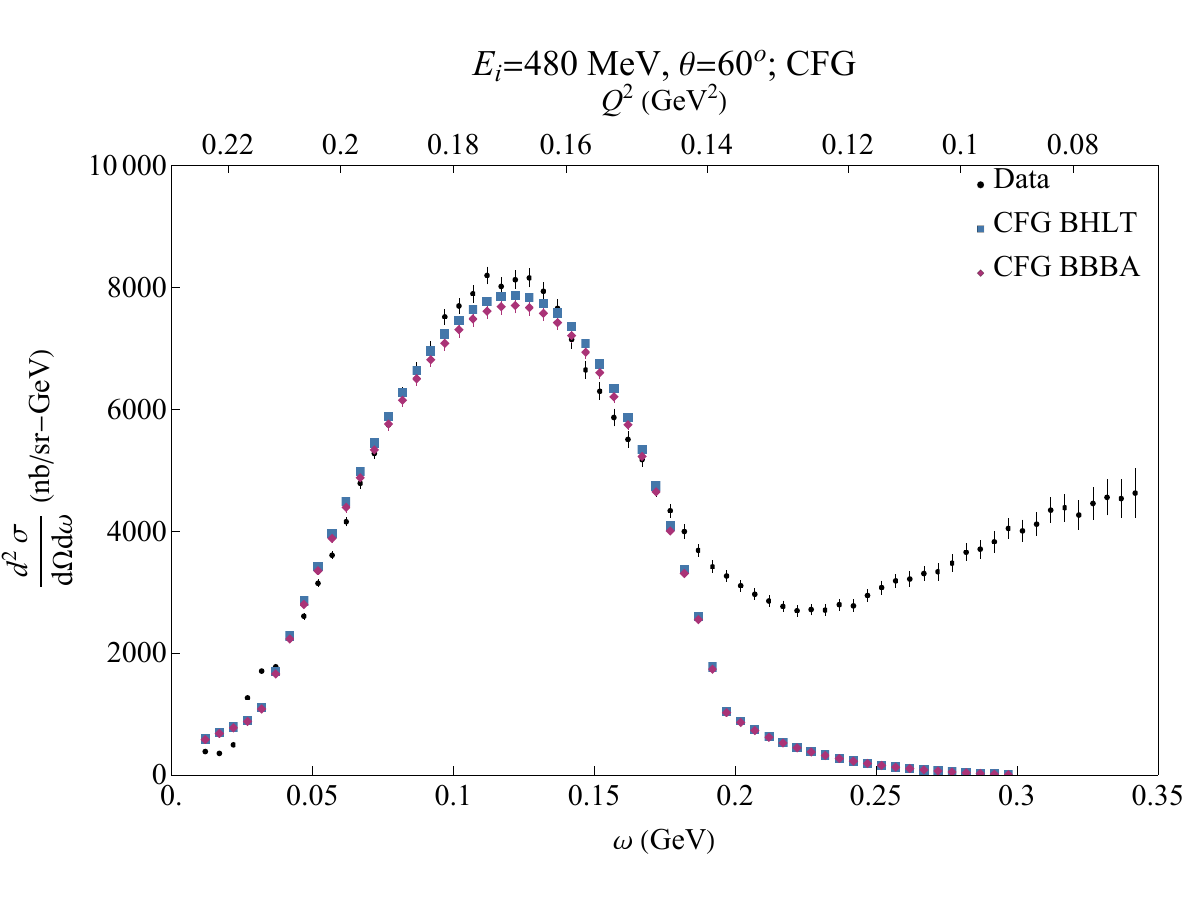}
\caption{\label{Fig:electron_FF_comparison} A comparison of RFG (left) and CFG (right) models to carbon data with BBBA \cite{Borah:2020gte} and BHLT  \cite{Borah:2020gte} parameterizations. The incoming electron energy is 480 MeV and the scattering angle is $60^\circ$.}
\end{center}
\end{figure}

\paragraph{Electron scattering:}
The analytic implementation of the CFG model allows to compare form factor and nuclear effects for electron-carbon scattering. 
First, we fix  
%vector form factors are fixed and the nuclear models are varied. Then, the nuclear model is fixed and the vector form factors are varied. 
%Fixing 
the vector form factors to the $z-$ expansion based parameterization of \cite{Borah:2020gte}, and compare the RFG and CFG models. An example of such a comparison is shown in figure \ref{Fig:680_MeV}  where there are clear differences between the two nuclear models. The same is true for other kinematical points, though the agreement with the data varies \cite{Bhattacharya:2024win}. Next, we fix the nuclear model and compare the vector form factors of \cite{Bradford:2006yz} and \cite{Borah:2020gte}. Performing the comparison separately for the RFG and CFG models, as shown in figure \ref{Fig:electron_FF_comparison}, the differences between form factors are small compared to  differences between nuclear models.  We conclude that for electron scattering the nuclear model differences are large, but form factor differences are small. 
%What happens for neutrino scattering?

\begin{figure}
\begin{center}
\includegraphics[scale=0.35]{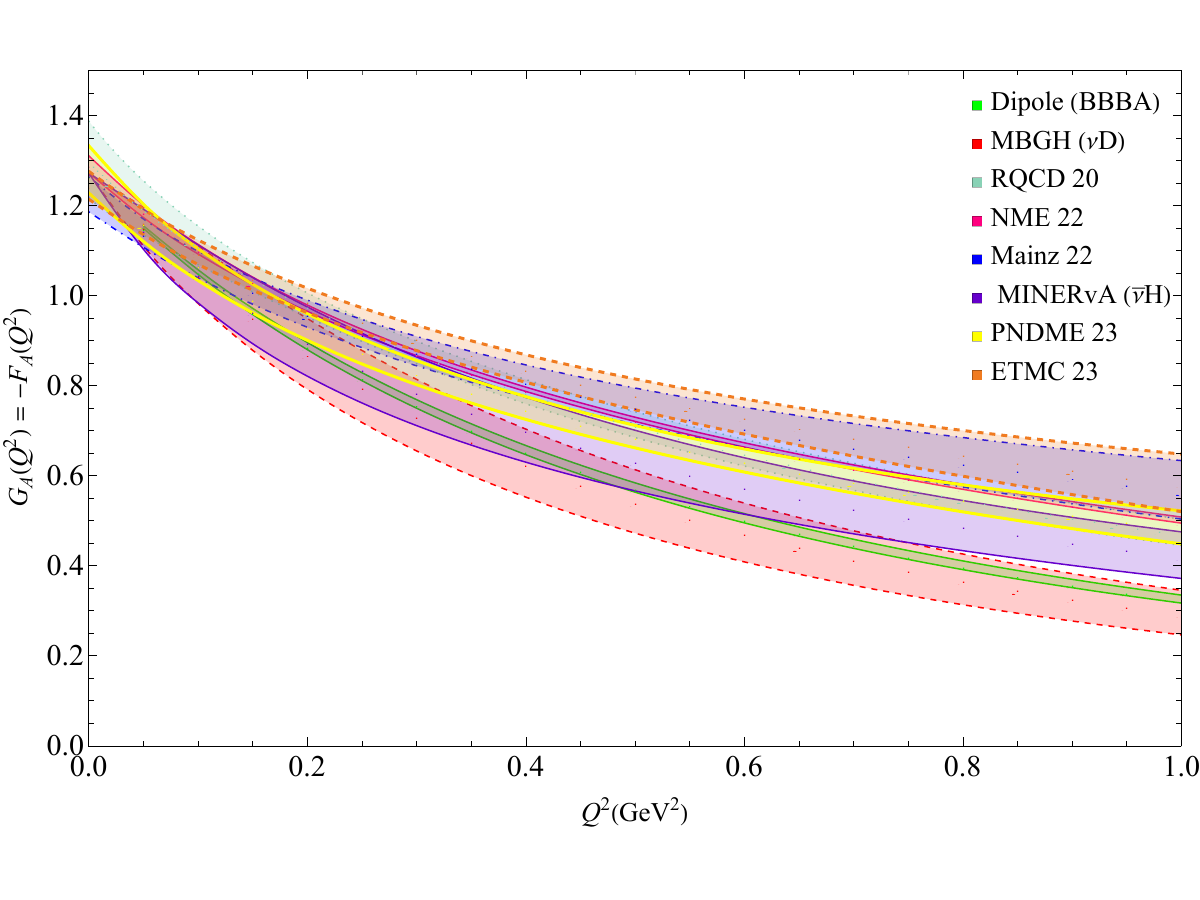}
\caption{\label{Fig:FA_Cmprsn} Comparison between the different axial form factor parameterizations implemented for neutrino scattering: Dipole (BBBA) using $M_A = 1.014\pm0.014$ GeV, MBGH \cite{Meyer:2016oeg}, RQCD 20 \cite{RQCD:2019jai}, NME 22 \cite{Park:2021ypf}, Mainz 22 \cite{Djukanovic:2022wru}, MINERvA \cite{MINERvA:2023avz}, PNDME 23 \cite{Jang:2023zts} and ETMC 23 \cite{Alexandrou:2023qbg}.}
\end{center}
\end{figure}

\paragraph{Neutrino scattering:}
%\paragraph{The axial form factor:}
Neutrino interaction involves also the axial current and not just the vector current.  This requires information about the axial form factor that is not accessible from electron scattering. Historically a one-parameter dipole model for the axial form factor was used. Such a model is not motivated by analyticity and is not flexible enough to describe the data. Instead, a systematic parameterization based on the $z$-expansion was suggested in \cite{Bhattacharya:2011ah}. In the last few years many such $z$-expansion based extractions of the axial form factor became available. Both from data \cite{Meyer:2016oeg, MINERvA:2023avz} and from lattice QCD \cite{RQCD:2019jai,Park:2021ypf,Djukanovic:2022wru, Jang:2023zts, Alexandrou:2023qbg}. Figure \ref{Fig:FA_Cmprsn} shows a comparison of these axial form factor parameterizations. It is clear that currently there is a large uncertainty on the axial form factor. 
%How does this uncertainty affect the theoretical predictions comparison to neutrino scattering data? 

\begin{figure}
\begin{center}
\includegraphics[scale=0.35]{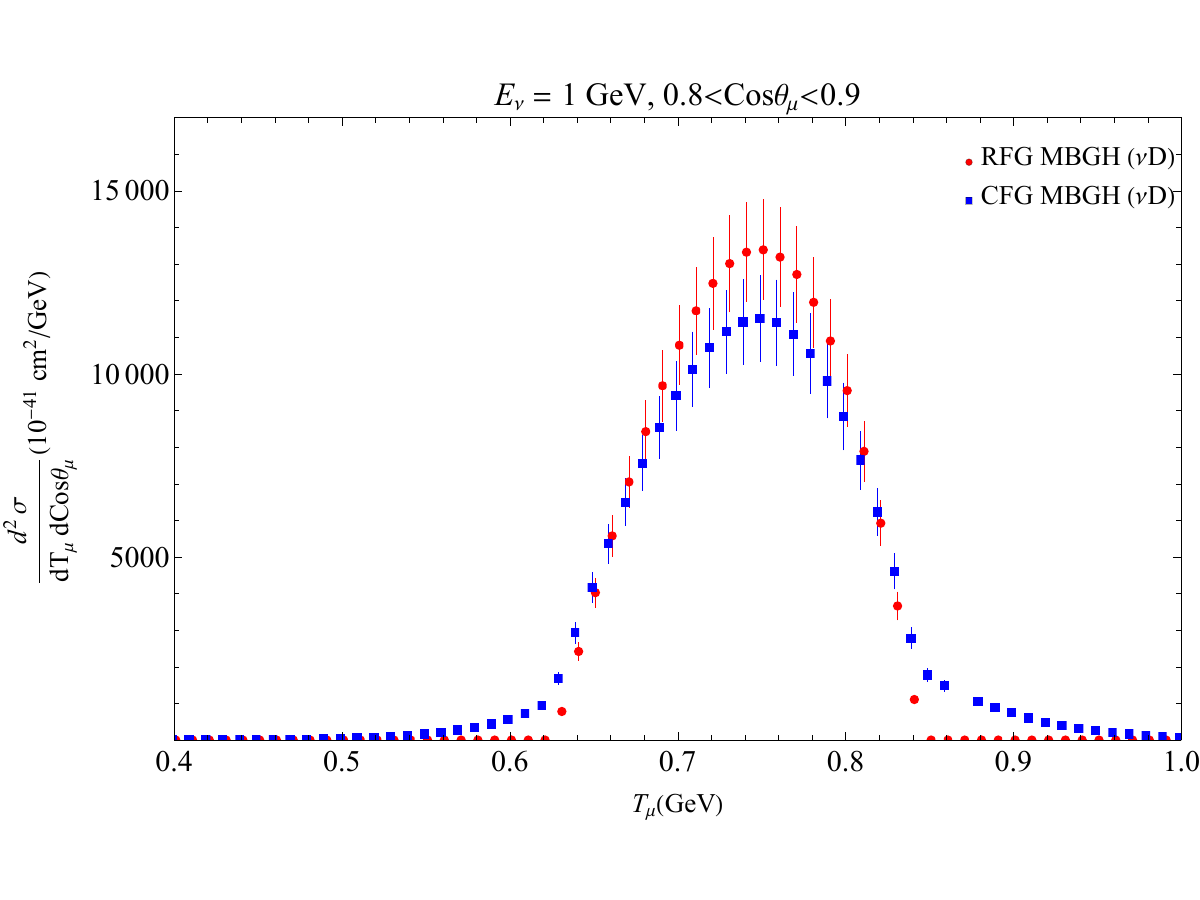}
\hspace{0.5cm}
\includegraphics[scale=0.35]{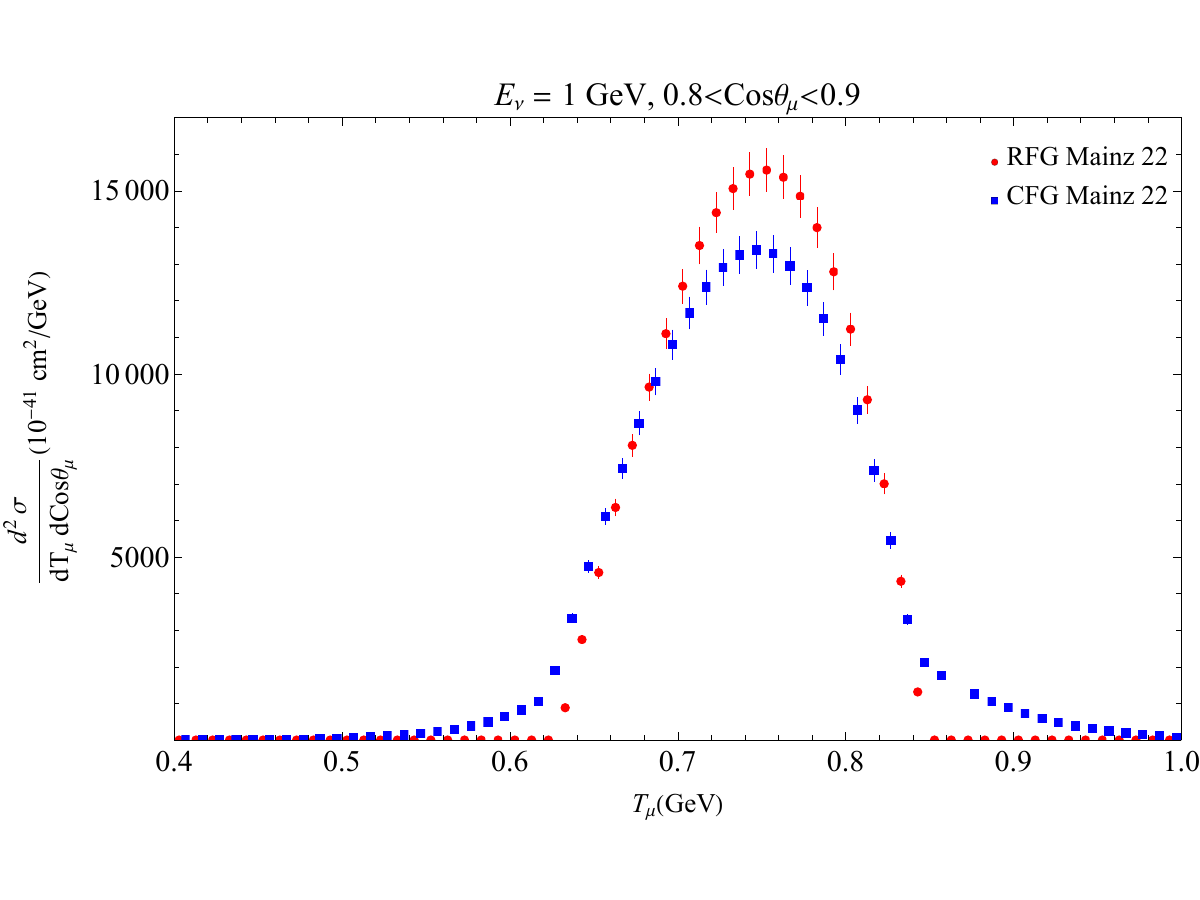}
\caption{\label{Fig:Fixed_Energy0.85} A comparison of  RFG (red circles) and CFG (blue squares) model double differential cross-section with incoming neutrino energy of $1$ GeV as a function of the outgoing muon kinetic energy. The scattering angle is $\cos\theta_\mu = 0.85$. Left: MBGH axial form factor \cite{Meyer:2016oeg} and Right: Mainz22 axial form factor  \cite{Djukanovic:2022wru}.}
\end{center}
\end{figure}

%\paragraph{Neutrino cross-section before flux averaging:}
We consider first  the hypothetical scattering of a mono-energetic 1 GeV neutrino on Carbon. We use the vector form factors of \cite{Borah:2020gte}. For clarity, we use the  two ``extremal" extractions of the axial form factor: MBGH \cite{Meyer:2016oeg} on the ``low" side and Mainz 22 \cite{Djukanovic:2022wru} on the ``high" side, see figure \ref{Fig:FA_Cmprsn}. Figure \ref{Fig:Fixed_Energy0.85} shows the comparison the two nuclear models and the two axial form extractions for the double differential scattering cross-section of $1$ GeV neutrino as a function of the outgoing muon kinetic energy $T_\mu$. The scattering angle is $\cos\theta_\mu = 0.85$.  One observes both differences between the two nuclear models and the two axial form factors extractions. 
%A mono-energetic neutrino is, generally speaking, a hypothetical situation. \\
%What happens after averaging over the neutrino energy flux? 

\begin{figure}
\begin{center}
\includegraphics[scale=0.35]{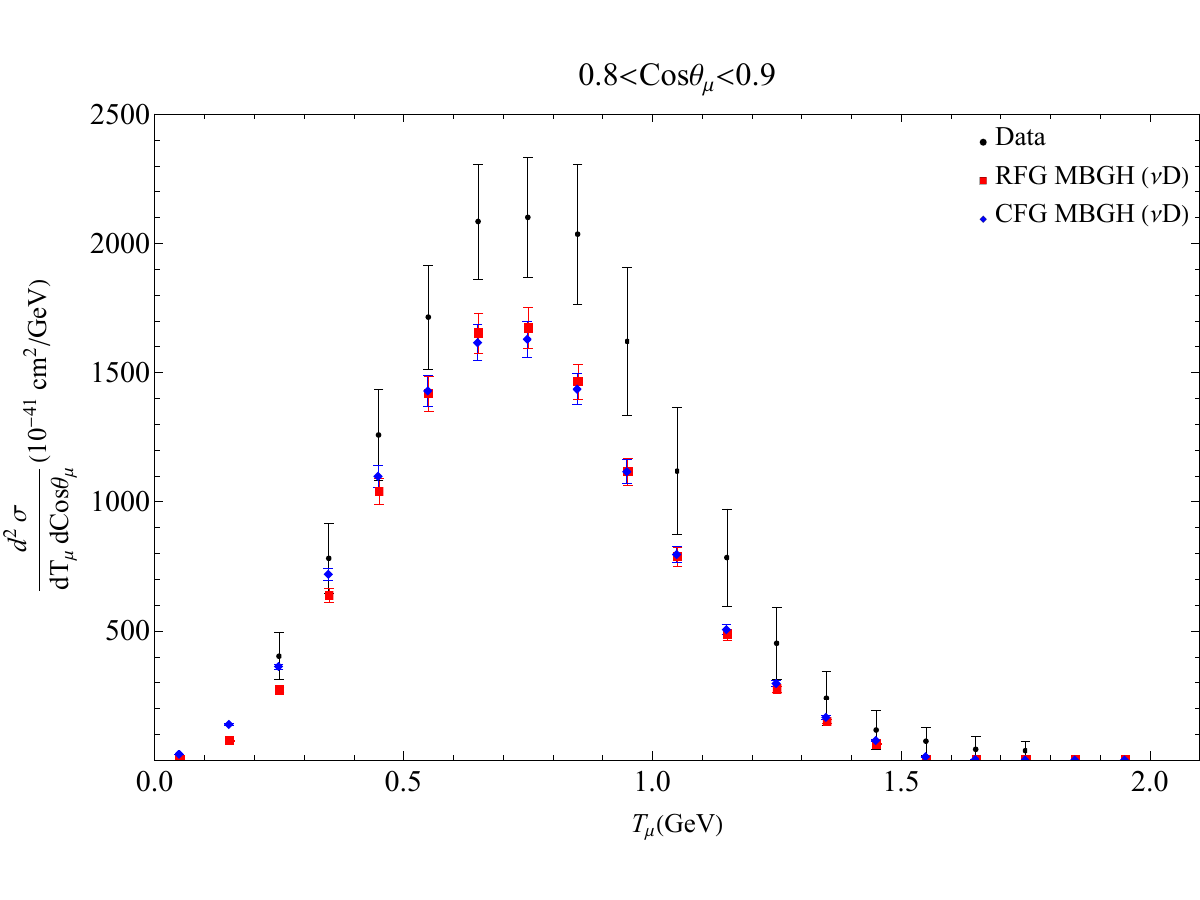}
\hspace{0.5cm}
\includegraphics[scale=0.35]{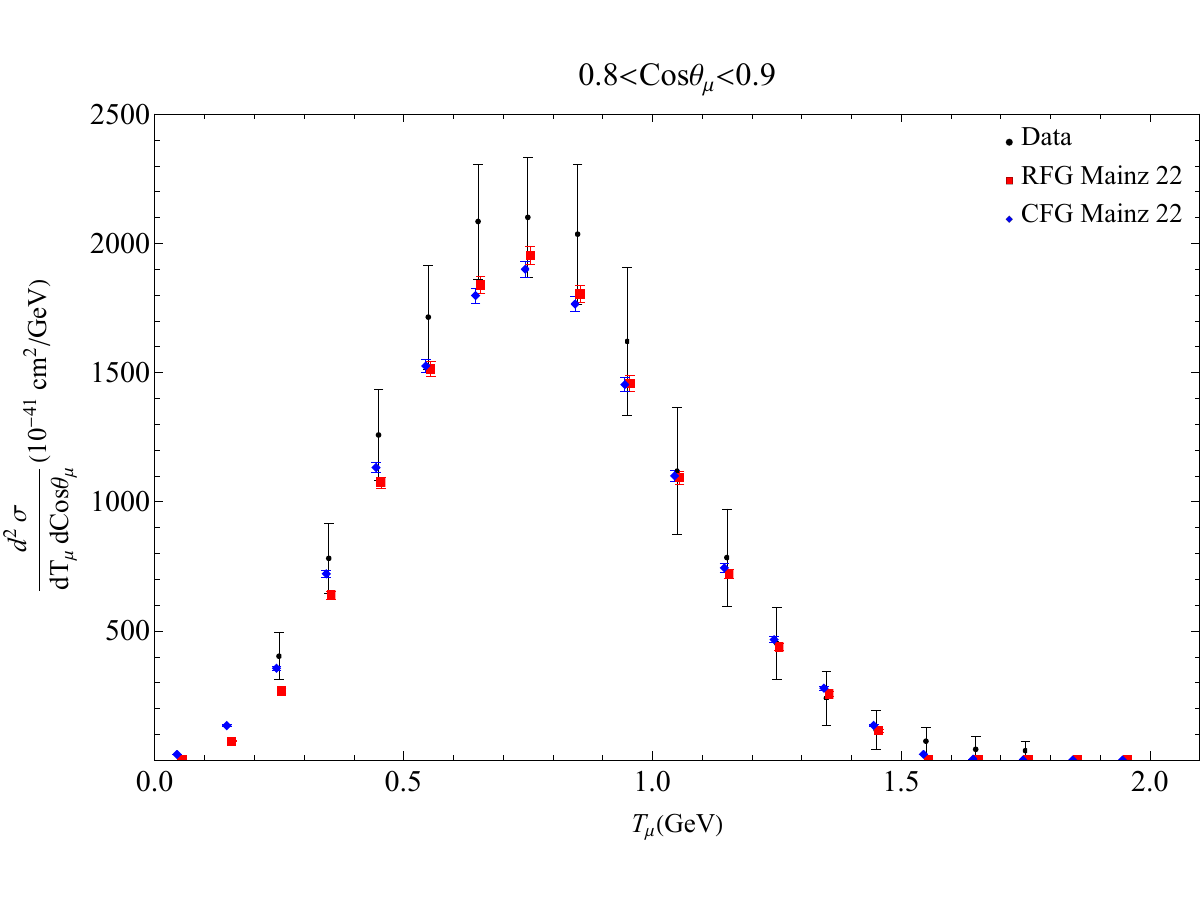}
\caption{\label{Fig:Theta_0.85} A comparison of RFG (red squares) and CFG (blue diamonds) models to flux averaged neutrino - carbon scattering data as a function of the outgoing muon kinetic energy scattering at an angle of $0.8 < \cos\theta_\mu < 0.9$ using Left: MBGH axial form factor Right: Mainz 22 for the axial form factor.}
\end{center}
\end{figure}

\begin{figure}
\begin{center}
\includegraphics[scale=0.35]{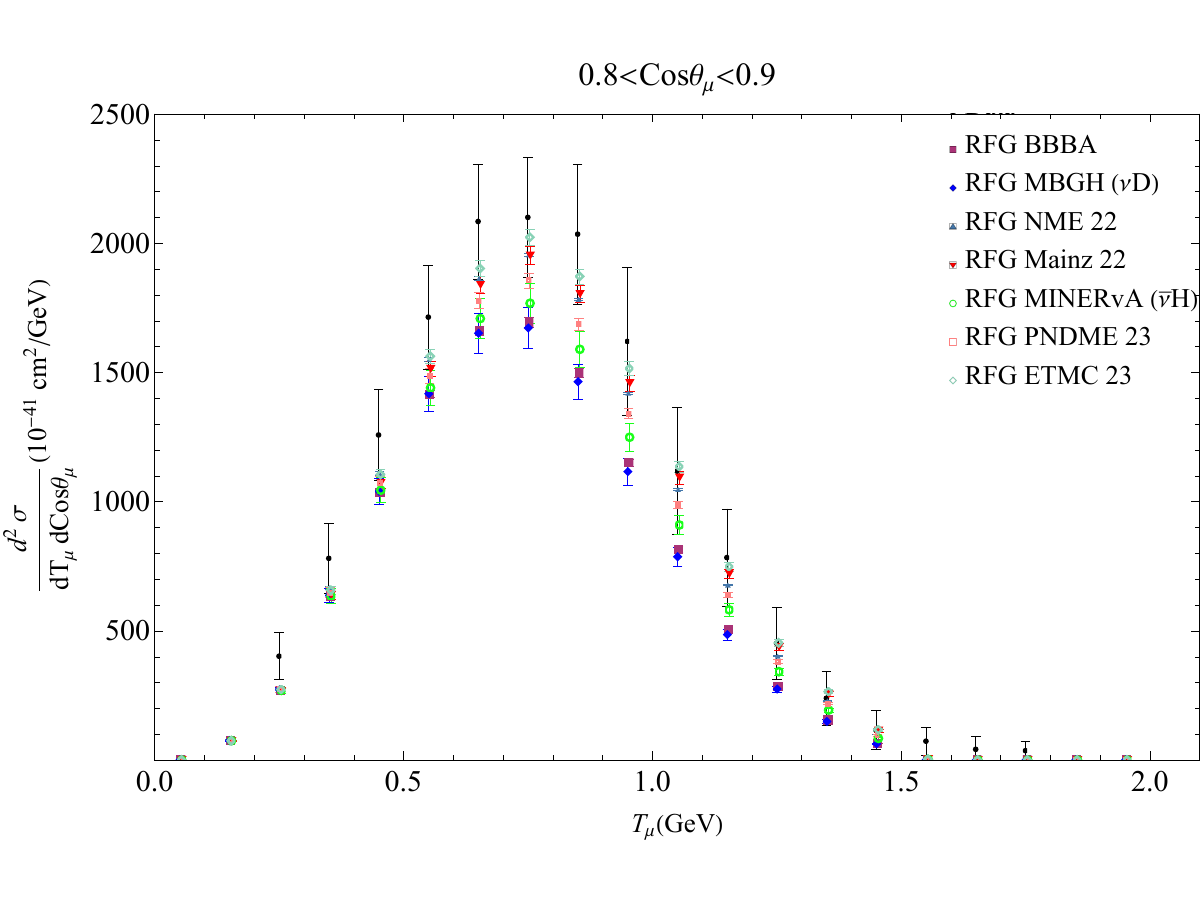}
\hspace{0.5cm}
\includegraphics[scale=0.35]{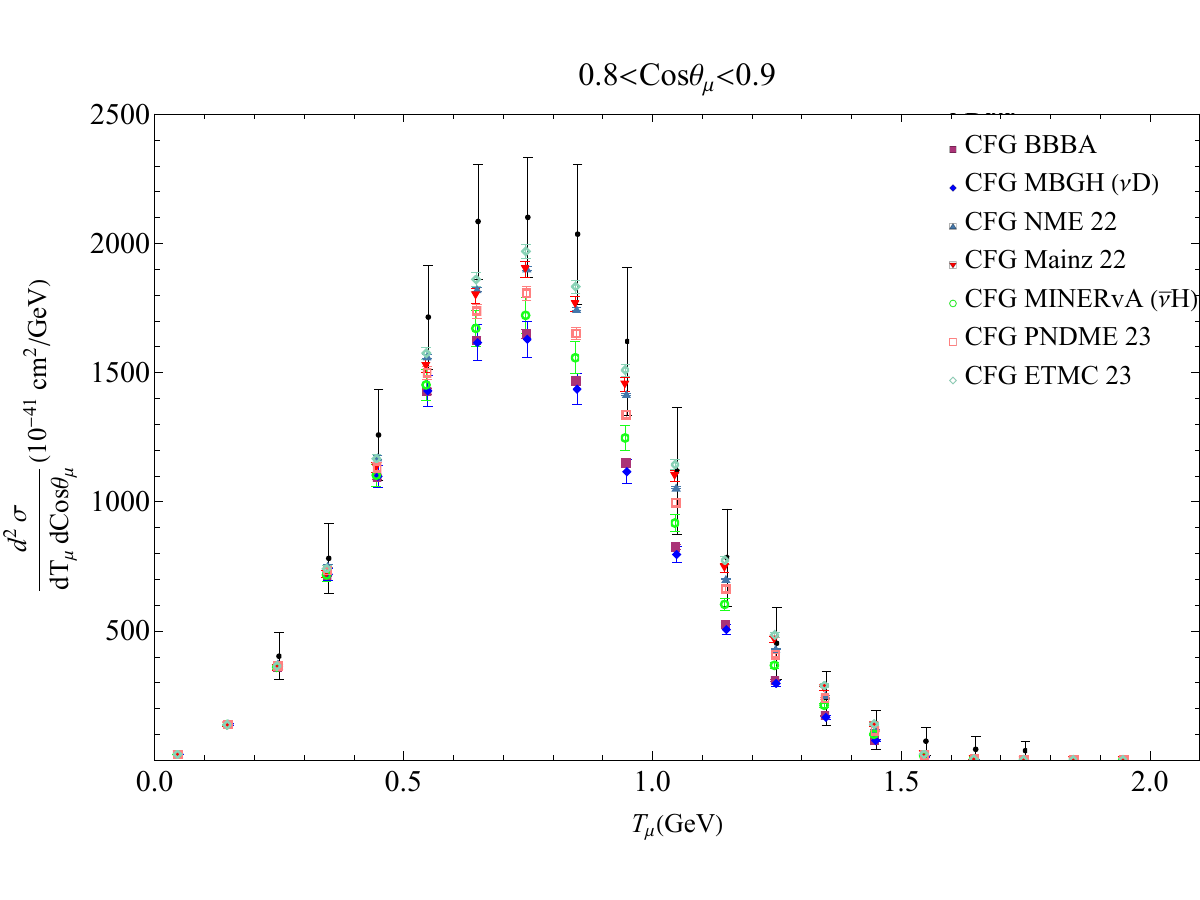}
\caption{\label{Fig:Theta_0.85_RFG_CFG} A comparison of BBBA, MBGH, NME 22, Mainz 22, MINERvA ($\bar\nu$H), PNDME 23 and ETMC 23  form factor parametrization to flux averaged neutrino - carbon scattering data as a function of the outgoing muon kinetic energy scattering at an angle of $0.8 < \cos\theta_\mu < 0.9$ using Left: RFG model Right: CFG model.}
\end{center}
\end{figure}

Next we compare nuclear models and form factor effects for the published MiniBooNE data \cite{MiniBooNE:2010bsu} by convoluting the cross section prediction with the published MiniBooNE muon-neutrino flux \cite{MiniBooNE:2010bsu}. First, we fix the axial form factor and vary the nuclear models and compare to the MiniBooNE data. The results are shown in figure \ref{Fig:Theta_0.85}. We can clearly see differences between the two axial form factor extractions  \cite{Meyer:2016oeg, Djukanovic:2022wru}, but after flux averaging, the RFG and CFG models are indistinguishable.
Next we fix the nuclear model and vary the axial form factor extractions shown in figure \ref{Fig:FA_Cmprsn}. The results are shown in figure \ref{Fig:Theta_0.85_RFG_CFG}. Again the RFG and CFG models are indistinguishable, but there is a continuous spread among  the axial form factor extractions.

\paragraph{Conclusions:} Current and future neutrino experiments require better control of systematic uncertainties in lepton-nucleus interactions.
In this talk we presented an analytic implementation of the CFG Model for lepton nucleus scattering \cite{Bhattacharya:2024win}. We then use the CFG and RFG models  to separate form factors and nuclear effects. For electron-nucleus scattering form factors differences are small, but nuclear models differences are large. For flux-averaged neutrino-nucleus scattering form factors differences are large, but nuclear models differences are small. Similar studies for other nuclear models would be welcomed, see, e.g., \cite{Simons:2022ltq}. 
%compared to data the Green's function Monte Carlo method and the Spectral Function formalism with 
%for a subset of the axial form factors of figure \ref{Fig:FA_Cmprsn}. 
We conclude with possible future directions for the CFG model. First, 
%after neutrino-flux averaging the CFG and RFG models were indistinguishable. C
can we separate CFG and RFG model for semi-inclusive neutrino-nucleus scattering?  Second, how to combine the CFG model with other effects, e.g., final state interaction (FSI). 

\paragraph{Acknowledgements:} This work was supported by the U.S. Department of Energy grant DE-SC0007983.

\end{document}